\documentclass[10pt,twocolumn]{IEEEtran}
\usepackage{epsfig,latexsym,subfigure,amsmath,cite}
\def\cT{\mathcal T}
\begin{document}

\title{Using the Physical Layer for Wireless Authentication in Time-Variant Channels}
\author{Liang~Xiao,~\IEEEmembership{Student~Member,~IEEE,}
        Larry~J.~Greenstein,~\IEEEmembership{Life~Fellow,~IEEE,}
        Narayan~B.~Mandayam,~\IEEEmembership{Senior~Member,~IEEE}
        and~Wade~Trappe,~\IEEEmembership{Member,~IEEE,}
        \thanks{The authors are with WINLAB, the Department
of Electrical and Computer Engineering, Rutgers University, North
Brunswick, NJ, 08902 USA e-mail: \{lxiao,
ljg,narayan,trappe\}@winlab.rutgers.edu}
\thanks{This research is supported, in part, through a grant CNS-0626439 from
the National Science Foundation.}
\thanks{Manuscript received February 2007; revised August 2007.}}
\maketitle
\begin{abstract}
The wireless medium contains domain-specific information that can
be used to complement and enhance traditional security mechanisms.
In this paper we propose ways to exploit the spatial variability
of the radio channel response in a rich scattering environment, as
is typical of indoor environments. Specifically, we describe a
physical-layer authentication algorithm that utilizes channel
probing and hypothesis testing to determine whether current and
prior communication attempts are made by the same transmit
terminal. In this way, legitimate users can be reliably
authenticated and false users can be reliably detected. We analyze
the ability of a receiver to discriminate between transmitters
(users) according to their channel frequency responses. This work
is based on a generalized channel response with both spatial and
temporal variability, and considers correlations among the time,
frequency and spatial domains. Simulation results, using the
ray-tracing tool WiSE to generate the time-averaged response,
verify the efficacy of the approach under realistic channel
conditions, as well as its capability to work under unknown
channel variations.
\end{abstract}

\section{Introduction}\label{Sec_intro}
As wireless devices become increasingly pervasive and essential,
they are becoming both a target for attack and the very weapon
with which such an attack can be carried out. Traditional
high-level computer and network security techniques can, and must,
play an important role in combating such attacks, but the wireless
environment presents both the means and the opportunity for new
forms of intrusion. The devices that comprise a wireless network
are low-cost commodity items that are easily available to
potential intruders and also easily modifiable for such intrusion.
In particular, wireless networks are open to intrusion from the
outside without the need for a physical connection and, as a
result, techniques that would provide a high level of security in
a wired network have proven inadequate in a wireless network, as
many motivated groups of students have readily
demonstrated\cite{Borisov:802.11,Arbaugh:Clothes, Walker:WEP}.

Although conventional cryptographic security mechanisms are
essential to securing wireless networks, these techniques do not
directly leverage the unique properties of the wireless domain to
address security threats. The physical properties of the wireless
medium are a powerful source of domain-specific information that
can be used to complement and enhance traditional security
mechanisms. In this paper, we propose a cross-layer approach to
augment the security of wireless networks for indoor wireless
environments. In particular, we believe that the nature of the
wireless medium can be turned to the advantage of the network
engineer when trying to secure wireless communications. The
enabling factor in our approach is that, in the rich multipath
environment typical of wireless scenarios, the response of the
medium along any transmit-receive path is {\em
frequency-selective} (or in the time domain, {\em dispersive}) in
a way that is {\em location-specific}. This means:
\begin{enumerate}
\item The channel can be specified by a number of complex samples
either in the frequency domain (a set of complex gains at a set of
frequencies) or the time domain (a set of impulse response samples
at a set of time delays). \item Such sets of numbers decorrelate
from one transmit-receive path to another if the paths are
separated by the order of an RF wavelength or more.
\end{enumerate} \noindent

While using the physical layer to enhance security might seem to
be a radical paradigm shift for wireless systems, we note that
this is not the first time that multipath and advanced physical
layer methods have proven advantageous. Specifically, we are
encouraged in our belief by two notable parallel paradigm shifts
in wireless systems: (1) code division multiple access (CDMA)
systems \cite{Book:CDMAVITERBI}, where the use of Rake processing
transforms multipath into a diversity-enhancing benefit; and (2)
multiple-input multiple-output (MIMO) antenna techniques
\cite{MIMO:original}, which transform scatter-induced Rayleigh
fading into a capacity-enhancing benefit.

Note that there have been recent efforts in studying the
information and secrecy capacity
\cite{TSE:channel_identification,hero_secure,negi_security}, that
can be achieved by using the radio channel information. In
contrast, this paper studies the feasibility of using such radio
channel information. It does so by explicitly devising hypothesis
testing procedures to estimate and track the radio channel for
authentication purposes.

We begin (Section \ref{sec:relatedworks}) by reviewing some
related work. Then (Section \ref{sec:Overview}), we provide an
overview of our proposed PHY-layer authentication service. We next
present a general time-variant channel model (Section
\ref{sec_channel}) that we will use as the basis for our
discussions in this paper. In Section \ref{sec_test}, we describe
a hypothesis testing framework for physical layer authentication.
In Section \ref{sec_sim}, we present an overview of our simulation
approach. We present our simulation results in Section
\ref{sec_results}, and wrap up the paper in Section
\ref{sec_conclusion} with concluding remarks.

\section{Related Work} \label{sec:relatedworks}
In commodity networks, such as 802.11 networks, it is easy for a
device to alter its MAC address and claim to be another device by
simply issuing an {\tt ifconfig} command. This weakness is a
serious threat, and there are numerous attacks, ranging from
session hijacking\cite{Arbaugh:Analysis} to attacks on access
control lists\cite{Arbaugh:Clothes}, which are facilitated by the
fact that an adversarial device may masquerade as another device.
In response, researchers have proposed using physical layer
information to enhance wireless security. For example, spectral
analysis has been used to identify the type of wireless network
interface card (NIC), and thus to discriminate among users with
different NICs \cite{XPassive_NIC_id:copeland}. A similar method,
radio frequency fingerprinting, discriminates wireless devices
according to the transient behavior of their transmitted signals
\cite{Xtransient:kranakis}. For more general networks, the clock
skew characteristic of devices has been viewed as a remote
fingerprint of devices over the Internet
\cite{Xremote_fingerprint:claffy}. In addition, the inherent
variability in the construction of various digital devices has
been used to detect intrusion \cite{Xsignal_fingerprintd:daniels}.

More recently, the wireless channel has been explored as a new
form of fingerprint for wireless security. The reciprocity and
rich multipath of the ultrawideband channel has been used as a
means to establish encryption keys
\cite{TSE:channel_identification}. In \cite{Stanford:Daniel}, a
practical scheme to discriminate between transmitters was proposed
and identifies mobile devices by tracking measurements of signal
strength from multiple access points. A similar approach was
considered for sensor networks in \cite{RSSI:Demirbas}. Concurrent
to these efforts, the present authors have built a significance
test that exploits the spatial variability of propagation to
enhance the authentication in the stationary, time-invariant
channel \cite{Liang_security:ICC}. In this paper, we have
significantly expanded the method to cover a more generalized
channel, where there are time variations due to changes in the
environment. As in \cite{Liang_security:ICC}, however, the ends of
the link remain stationary, as might be the case for a population
of users sitting in a room or airport terminal. We will see that,
in some cases, the time variations \textit{improve} the
authentication.

\section{Problem Overview} \label{sec:Overview}
Authentication is traditionally associated with the assurance that
a communication comes from a specific entity\cite{Wadebook}. In
the context of physical layer authentication, however, we are not
interested in identity, \textit{per se}, but rather are interested
in recognizing a particular transmitting device. The ability to
distinguish between different transmitters would not replace
traditional identity-based authentication, but would be
particularly valuable as a wireless system enhancement. Such an
approach would be beneficial for scenarios where managing
cryptographic key material is difficult, and further would reduce
the load placed on higher-layer authentication buffers.

\begin{figure}[t]
\begin{center}
       \epsfig{figure=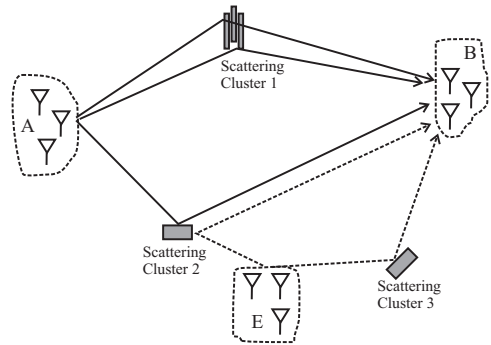,width=2.5in}
\end{center}
\vspace{-3mm}\caption{\label{fig:SEVILLE} The adversarial
multipath environment involving multiple scattering surfaces. The
transmission from Alice (A) to Bob (B) experiences different
multipath effects than the transmission by the adversary, Eve
(E).}
\end{figure}

Here, we borrow from the conventional terminology of the security
community by introducing three different parties: Alice, Bob and
Eve. For our purposes, these three entities may be thought of as
wireless transmitters/receivers that are potentially located in
spatially separated positions, as depicted in Fig.
\ref{fig:SEVILLE}. Our two ``legal" protagonists are the usual
Alice and Bob, and for the sake of discussion throughout this
paper, Alice will serve as the transmitter that initiates
communication, while Bob will serve as the intended receiver.
Their nefarious adversary, Eve, will serve as an active adversary
that injects undesirable communications into the medium in the
hopes of spoofing Alice.

Our security objective, broadly speaking, is to provide
authentication between Alice and Bob, despite the presence of Eve.
Since Eve is within range of Alice and Bob, and capable of
injecting her own signals into the environment to impersonate
Alice, Bob must have the ability to differentiate between
legitimate signals from Alice and illegitimate signals from Eve.

Consider a simple transmitter identification protocol in which Bob
seeks to verify that Alice is the transmitter. Suppose that Alice
transmits probes into the channel at a rate sufficient to assure
temporal coherence between channel estimates and that, prior to
Eve's arrival, Bob has estimated the Alice-Bob channel. Eve wishes
to convince Bob that she is Alice. Bob will require that each
information-carrying transmission be accompanied by an
authenticator signal. The channel response to a transmitted signal
between Alice and Bob is a result of the multipath environment.
Bob may use the received version of the authenticator signal to
estimate the channel response and compare this with a previous
record for the Alice-Bob channel. If the two channel estimates are
``close" to each other, then Bob will conclude that the source of
the message is the same as the source of the previously sent
message. If the channel estimates are not similar, then Bob should
conclude that the source is likely not Alice.

There are several important issues related to such a procedure
that should be addressed before it can be a viable authentication
mechanism. First is the specification of the authenticator signal
that is used to probe the channel. There are many standardized
techniques to probe the channel, ranging from pulse-style probing
(including PN sequences) to multi-tonal probing
\cite{RappaportBook}, and we may use these techniques to estimate
the channel response. Regardless of what probing method is
employed, the channel response can be characterized in the
frequency domain, and throughout this paper we will represent our
channels in that domain.

A second issue is that in a richly scattered multipath environment
(typical of indoor wireless environments), it is difficult for an
adversary to create or precisely model a waveform that is
transmitted and received by entities that are more than a
wavelength away from the adversary. This assertion is supported by
the well-known Jakes uniform scattering model \cite{Jakes}, which
states that the received signal rapidly decorrelates over a
distance of roughly half a wavelength, and that spatial separation
of one to two wavelengths is sufficient for assuming independent
fading paths. The implication of such a scattering model in a
transmitter identification application remains to be tested, and a
key objective of this study is to examine the utility of a typical
indoor multipath environment for discriminating between Alice-Bob
and Eve-Bob channels.

Finally, it should also be noted that the channel response may
change with time due to changes in the environment (people moving,
doors opening or closing) and in practice it will be necessary to
guarantee the continuity of the authentication procedure by
probing the channel at time intervals less than the channel's
``coherence time". This paper examines the ability to authenticate
transmitters in such a time-variant environment, and serves to
illustrate the potential for new forms of physical layer security.
\section{Channel Model}\label{sec_channel}
\subsection{Basic Form}
We assume that Bob first measures and stores the frequency
response of the channel connecting Alice with him. Due to his
receiver thermal noise, Bob stores a noisy version of the channel
response, $H_{A}(f)$. After awhile, he has to decide whether a
transmitting terminal is still Alice, based on a noisy measured
version, $H_t(f)$, of that terminal's channel response to Bob. By
sampling $H_{A}(f)$ and ${H}_{t}(f)$ at $f\in (f_o-W/2,f_o+W/2]$,
Bob obtains two frequency response vectors, ${\textbf H}_{A}$ and
${\textbf H}_{t}$, of length $M$, where $W$ is the measurement
bandwidth; $f_o$ is the center frequency of the measurement; and
the vector elements are frequency response samples at $M$
uniformly spaced frequencies over the measurement bandwidth.

We consider a generalized time-variant channel response, where
each frequency response sample is made up of three parts: the
fixed part that is the average channel response over time and
contains the spatial variability information, the variable part
with zero mean, and the receiver noise. Thus the $m$-th element of
${\textbf H}_{A}$ at time $kT$ from some arbitrary time origin can
be written as
\begin{align}
H_{A,m}[k]=\overline{H}_{A,m}+\epsilon_{A,m}[k]+N_{A,m}[k],\quad
1\leq m \leq M, \label{eq_chA}
\end{align}
where we use the notation that $X_m[k]$ is the sample from
$X(t;f)$ at the $m$-th tone at a sampling time of $kT$. More
specifically, $X_m[k]=X(kT;f_o-W/2+m \Delta f)$, $m=1,\cdots,M$,
where $\Delta f=W/M$; $M$ is the sample size in the frequency
domain; and $T$ is the sampling interval. The term
$\overline{H}_{A,m}$ is the average value of the $m$-th tone over
time, $\epsilon_{A,m}[k]$ is the zero-mean variable part at time
$kT$, and $N_{A,m}[k]$ represents thermal noise sample at the
$m$-th tone at time $kT$. The noises are modelled by
$CN(0,\sigma_N ^2)$, i.e., zero-mean complex Gaussian samples with
variance $\sigma_N ^2$. Without loss of realism, we can assume
that they are independent across time, tone (frequency) and
terminal (space), and that $\epsilon_{A,m}[k]$ is independent of
$N_{A,m}[k]$.

\subsection{Delay Profile and Doppler Spectrum (Temporal Fading) of the Variable Part}
We model the variable part of the channel response as
\textit{wide-sense stationary uncorrelated scattering} (WSSUS),
and can thus use a multipath tapped delay line to model its
impulse response, $h(t,\tau)$ \cite{bello:WSSUS}:
\begin{align}
h(t,\tau)=\sum_{l=0}^{\infty} A_l(t)\delta(t-l\Delta \tau),
\end{align}
where $t$ is the observation time, and $l\Delta \tau$ and $A_l(t)$
are, respectively, the delay and complex amplitude of the $l$-th
multipath component, with $E[A_l(t)]=0$ over time. We set $\Delta
\tau=1/W$, since the receiver cannot resolve two components with
time difference smaller than the inverse of the bandwidth.

The frequency response of the variable part is the Fourier
transform of $h(t;\tau)$ in terms of $\tau$,
\begin{align}
\epsilon_{A,m}[k]&={\cal F} \{h(t;\tau)\}|_{t=kT,f=f_o-W/2+m
\Delta f}\nonumber\\&=\sum_{l=0}^{\infty} A_l[k]e^{-j2\pi
(f_o-W/2+m \Delta f)l \Delta}, \label{e_epsil}
\end{align}
where $A_l[k]=A_l(kT)$ is the amplitude sample of the multipath
component at time $kT$.

For illustrative purposes, we use the one-sided exponential
distribution to model the power delay spectrum of
$A_l[k]$\footnote{The literature abounds with empirical data
\cite{Greenstein:99_04} and theoretical examples
\cite{bello:theoretic} in which the exponential delay profile
appears. We invoke it here for the sake of concreteness, which
will allow us to compute numerical results, but we also recognize
it to be a realistic condition.}, i.e.,
\begin{align}
P_\tau [l]=\textrm{Var}[A_l[k]]=\sigma
_T^2(1-e^{-\gamma\Delta\tau})e^{- \gamma \Delta \tau l},
\label{e_var_al}
\end{align}
where $\gamma=2\pi B_c$ is the inverse of the average delay
spread, $B_c$ is the coherence bandwidth of the variable part, and
$\sigma _T^2$ is the average power of $A_l[k]$ over all taps.

Also for illustrative purposes, we use an autoregressive model of
order 1 (AR-1) to characterize the temporal process of $A_l[k]$,
i.e.,
\begin{align}
A_l[k]=a A_l[k-1]+\sqrt{(1-a^2)P_\tau[l]}u_l[k], \label{e_at}
\end{align}
where the AR coefficient $a$ denotes the similarity of two $A_l$
values spaced by $T$ and the random component $u_l[k] \sim
CN(0,1)$ is independent of $A_l[k-1]$.

\subsection{Spatial Correlations}
As pointed out in Section \ref{Sec_intro}, in a typically rich
scattering environment, the radio channel response decorrelates
quite rapidly in space. Later, we will cite the use of the
ray-tracing software WiSE to emulate the spatial correlation
characteristics of the \emph{fixed} part of the channel response
($\overline{H}$). As to the \emph{variable} part, however, we
consider the two extreme cases:

\noindent 1) \textit{Spatially independent and identically
distributed} $\epsilon$. The frequency response sample of the
channel between Eve and Bob can be written as
\begin{align}
H_{E,m}[k]=\overline{H}_{E,m}+\epsilon_{E,m}[k]+N_{E,m}[k],
\label{eq_chE1}
\end{align}
where $1\leq m \leq M$, $\overline{H}_{E,m}=E[H_{E,m}[k]]$ is the
time average; thermal noise $N_{E,m}[k]\sim CN(0,\sigma^2 _N)$;
and $\epsilon_{E,m}[k]$ and $\epsilon_{A,m}[k]$ are independent
identically distributed (i.i.d.).

\noindent 2) \textit{Complete spatially correlated variation}
($\epsilon_{E,m}[k]=\epsilon_{A,m}[k]$). Here, we have
\begin{align}
H_{E,m}[k]=\overline{H}_{E,m}+\epsilon_{A,m}[k]+N_{E,m}[k],\quad
1\leq m \leq M. \label{eq_chE2}
\end{align}

\subsection{Important Relationships}
Two important relationships we will use in the hypothesis testing
later are as follows (proofs are provided in the Appendix):

\textit{Relationship 1}: \begin{align} {\textbf H}
_{A}[k]-{\textbf H} _{A}[k-1] \sim CN(\underline{0},
\textbf{R}),\label{E_heha}
\end{align}
where

\begin{align}
\textbf{R}&=\textrm{Cov}[{\textbf H} _{A}[k]-{\textbf H}
_{A}[k-1]]\nonumber\\&=[r(m-n)] _{mn}, \quad 1 \leq m,n \leq M,
\label{E_R1}
\end{align}
$r(0)=2(1-a) \sigma _T ^2 +2 \sigma _N ^2$, and
\begin{align}
r(m)=\frac{2\sigma^2_T(1-a)(1-e^{-2\pi B_c/W})}{1-e^{-2\pi
B_c/W-j2\pi m/M}},\: 1-M \leq m \leq M-1. \label{eq_r_m}
\end{align}

\textit{Relationship 2}: For the case with spatially independent
time variation,
\begin{align}
{\textbf H} _{E}[k]-{\textbf H} _{A}[k-1] \sim
CN((\overline{{\textbf H}}_E-\overline{{\textbf H}}_A),
\textbf{G}),\label{E_heha}
\end{align}
where
\begin{align}
\textbf{G}&=\textrm{Cov}[{\textbf H} _{E}[k]-{\textbf H}
_{A}[k-1]]\nonumber\\&=\left [
\begin{array}{cccc} 2\sigma _T ^2 +2 \sigma _N ^2 &
\frac{r(-1)}{1-a} & \cdots & \frac{r(1-M)}{1-a}\\\frac{r(1)}{1-a}
& 2\sigma _T ^2 +2 \sigma _N
^2 & \cdots & \frac{r(2-M)}{1-a}\\\cdots & \cdots & \cdots & \cdots \\
\frac{r(M-1)}{1-a}&\frac{r(M-2)}{1-a}&\cdots&2 \sigma _T ^2 +2
\sigma _N ^2 \label{E_G1}
\end{array}  \right ].
\end{align}
\section{Hypothesis Testing}\label{sec_test}
Here, we present formulas for hypothesis testing that will be
reduced later to numerical results.
\subsection{General Case}
As in \cite{Liang_security:ICC}, Bob uses a simple hypothesis test
to decide if the transmitting terminal is Alice or a would-be
intruder, Eve. The null hypothesis, ${\cal H}_0$, is that the
terminal is not an intruder, i.e. the claimant is Alice; and Bob
accepts this hypothesis if the test statistic he computes, $Z$, is
below some threshold, $\mathcal{T}$. Otherwise, he accepts the
alternative hypothesis, ${\cal H}_1$, that the claimant terminal
is someone else. Thus,
\begin{align}
{\cal H}_0: \quad & \textbf{H}_{t}[k] = \textbf{H}_{A}[k]\label{test1}\\
{\cal H}_1: \quad & \textbf{H}_{t}[k] \neq \textbf{H}_{A}[k]
,\label{test2}
\end{align}

First, we assume spatially independent time variations and assume
Bob knows the key channel variation parameters $a$, $B_c$ and
$\sigma_T$. (We will discuss other cases in the later parts of
this section.) We choose the test statistic in this default
setting as
\begin{align}
Z=\textbf{z} ^{H} \textbf{z} =2( \textbf{H} _{t}[k]- \textbf{H}
_{A}[k-1])^H \textbf{R} ^{-1} (\textbf{H} _{t}[k]-\textbf{H}
_{A}[k-1]),\label{E_test_sta_1}
\end{align}
where \textbf{z} =$\sqrt{2}({\textbf R}_d^{H})^{-1}({\textbf H}
_{t}[k]-{\textbf H}_A[k-1])$, \textbf{R} and $\textbf R_d$ are the
covariance matrix of ${\textbf H}_{A}[k]-{\textbf H}_{A}[k-1]$,
(\ref{E_R1}), and its Cholesky factorization (i.e.,
\textbf{R}=\textbf{R}$_{d}^H$ \textbf{R}$_d$).

It can be shown that, when the transmitting terminal is Alice,
each element of \textbf{z} is i.i.d., following a normal
distribution, ${\textbf z}=\sqrt{2}({\textbf
R}_d^{H})^{-1}({\textbf H}_{A}[k]-{\textbf H}_{A}[k-1])$, where
the elements are i.i.d., and $z_i \sim CN(0,2)$, $1\leq i\leq M$.
Thus the test statistic $Z$ is a chi-square random variable with
$2M$ degrees of freedom \cite{Book:math}, i.e., $Z={\textbf
z}^{H}{\textbf z}\sim \chi ^2 _{2M}$.

We define the rejection region for ${\cal H}_0$ as
$Z>\mathcal{T}$. Thus, the ``false alarm rate" (or Type I error)
is $\alpha={\textrm P_r} \{Z>\mathcal{T}|{\cal
H}_0\}=1-\textrm{F}_{\chi^2_{2M}}(\mathcal{T})$; and the ``miss
rate" (or Type II error) is given by (\ref{E_beta_0}),
\begin{figure*}[!htp]
\normalsize 
\setcounter{equation}{15}
 \begin{equation}
\beta ={\textrm P_r} \{ Z<\mathcal{T} |{\cal{H}}_1\}={\textrm P_r}
\{2(\textbf{H}_{E}[k]-\textbf{H} _{A}[k-1])^H
\textbf{R}^{-1}({\textbf H}_{E,t}[k]-\textbf{H} _{A}[k-1])<
{\textrm F}^{-1}_{\chi^2_{2M}}(1-\alpha)\},\label{E_beta_0}
\end{equation}
\vspace*{4pt}
\end{figure*}
where $\textrm{F}_X(\cdot)$ is the CDF of the random variable $X$
and $\textrm{F}^{-1}_X(\cdot)$ is the inverse function of
${\textrm F}_X(\cdot)$. For a specified $\alpha$, the threshold of
the test is $\mathcal{T}=\textrm{F}^{-1}_{\chi^2_{2M}}(1-
\alpha)$, and the miss rate can be obtained by numerical methods.

\subsection{Asymptotic Results for Low Correlation Bandwidth}
When the variation is independent over tones (i.e., $B_c/W \ll
1$), the covariance matrices of Eq. (\ref{E_R1}) and (\ref{E_G1})
become
\begin{align}
\textbf{R}&=\textrm{Cov}[\textbf{H}_{A}[k]-\textbf{H} _{A}[k-1]]=(2(1-a) \sigma _T ^2 +2 \sigma _N ^2)\textbf{I}\nonumber\\
\textbf{G}&=\textrm{Cov}[\textbf{H}_{E}[k]-\textbf{H}
_{A}[k-1]]=(2\sigma _T ^2 +2 \sigma _N ^2)\textbf{I},
\end{align}
where \textbf{I} is the identity matrix. Thus the test statistic
Eq. (\ref{E_test_sta_1}) becomes
\begin{align}
Z=\frac{|\textbf{H}_{t}[k]-\textbf{H}_{A}[k-1]|^2}{(1-a) \sigma _T
^2 + \sigma _N ^2}=Z_2/\rho ,
\end{align}
where
\begin{align} \rho=\frac{(1-a) \sigma _T ^2 + \sigma _N
^2}{\sigma _T ^2 + \sigma _N ^2}. \label{E_rho}
\end{align}

It is easy to see that, under ${\cal H}_1$, the test statistic is
a non-central chi-square distribution with order $2M$, i.e., $ Z_2
\sim \chi^2_{2M,\mu}$ with non-central parameter
\begin{align} \mu
=\frac{\sum_{m=1}^M|\overline{H}_{E,m}-\overline{H}_{A,m}|^2}{\sigma
_T ^2 + \sigma _N ^2}\label{E_mu}
\end{align}
Thus, the miss rate for specified $\alpha$, (\ref{E_beta_0}), can
be written as
\begin{align}
\beta={\textrm P_r} \{Z<\mathcal{T}|{\cal
H}_1\}=\textrm{F}_{\chi^2_{2M,\mu}}(\rho
\textrm{F}^{-1}_{\chi^2_{2M}}(1-\alpha)). \label{E_beta_1}
\end{align}

\subsection{Asymptotic Results for High Correlation Bandwidth}
When the variation is totally correlated over tones (i.e., $B_c/W
\gg 1$), the covariance matrices of (\ref{E_R1}) and (\ref{E_G1})
degrade to
\begin{align}
\textbf{R}&=2 \sigma _N ^2\textbf{I}+2(1-a) \sigma _T ^2 \textbf{1} \\
\textbf{G}&=2 \sigma _N ^2\textbf{I}+2\sigma _T ^2 \textbf{1},
\end{align}
where \textbf{1} is a $M \times M$ matrix with each element equals
to 1. Again, we can use Eq. (\ref{E_beta_0}) to numerically
calculate the miss rate $\beta$ for specified false alarm rate
$\alpha$.

\subsection{Unknown Parameters}
When Bob does not know the parameters $a$, $B_c$ and $\sigma_T$,
it is reasonable for him to use as the test statistic
\begin{align}
Z=\frac{1}{\sigma _N ^2}|\textbf{H} _{t}[k]-\textbf{H}
_{A}[k-1]|^2.
\end{align}
In this case, we can obtain numerical results for the false alarm
rate and miss rate for specified threshold $\mathcal{T}$, plotting
$\beta$ vs. $\alpha$ with $\mathcal{T}$ as an implicit parameter.

\subsection{Full Spatial Correlation}
Now we consider the other extreme case of spatial correlation,
namely, $\epsilon_{E,m}[k]=\epsilon_{A,m}[k]$ (full spatial
correlation). The spatial correlation has no impact under the
hypothesis ${\cal H}_0$. However, under the hypothesis ${\cal
H}_1$, the correlation matrix of the difference between two
measurements becomes $\textbf{R}$, and ${\textbf
H}_{E}[k]-{\textbf H}_{A}[k-1] \sim CN((\overline{\textbf
H}_E-\overline{{\textbf H}}_A), \textbf{R})$. Thus, the test
statistic under ${\cal H}_1$ is non-central chi-square
distributed, $Z =|\sqrt{2}({\textbf R}_d^{H})^{-1}({\textbf H}
_{E}[k]-{\textbf H}_{A}[k-1])|^2 \sim \chi^2_{2M,\mu}$, with
non-central parameter $\mu =|\sqrt{2}( {\textbf R}_d^{H})^{-1}(
{\overline{\textbf H}}_E- {\overline{\textbf H}} _{A})|^2$.
Therefore, the miss rate for the fully spatially correlated
temporal variation can be written as
\begin{align}
\beta=\textrm{F}_{\chi^2_{2M,\mu}}(\textrm{F}^{-1}_{\chi^2_{2M}}(1-\alpha)).\label{E_beta_end}
\end{align}

\subsection{Discussion: Impact of Time Variations}\label{subsec:variation}
As a benchmark, from ({\ref{E_beta_1}}) we have the miss rate for
the time-invariant channel as \cite{Liang_security:ICC},
\begin{align}
\beta=\textrm{F}_{\chi^2_{2M,\mu}}(\textrm{F}^{-1}_{\chi^2_{2M}}(1-\alpha)),
\label{E_beta_2}
\end{align}
where
$\mu=\sum_{m=1}^M|\overline{H}_{E,m}-\overline{H}_{A,m}|^2/{\sigma
_N ^2}$.

In the presence of time variation, however, the miss rate may
become smaller. The asymptotic miss rate for the time-variant
channel at high bandwidth, (\ref{E_beta_1}), increases with
$\rho$, (\ref{E_rho}), and decreases with $\mu$, (\ref{E_mu}). As
the time variation $\sigma_T ^2$ rises from 0 to $\infty$, $\rho$
decreases from 1 to $1-a$ and $\mu$ falls from
$\sum_{m=1}^M|\overline{H}_{E,m}-\overline{H}_{A,m}|^2/\sigma _N
^2$ to 0, which may results in a smaller miss rate.

Actually, the temporal-variation has a two-fold impact: 1) It adds
uncertainty to the channel from Alice, and thus Bob has to
increase the test threshold to accept Alice (negative impact on
the performance); 2) the variation is usually strongly correlated
in time while very weakly correlated in space, and thus $\epsilon
_A[k]-\epsilon _A[k-1] < \epsilon _E[k]-\epsilon _A[k-1]$
(positive impact on performance).

When $\sigma _T$ is negligible, the channel can be viewed
approximately as a time-invariant one, wherein the miss rate is
given by (\ref{E_beta_2}). As $\sigma_ T$ rises, the miss rate
falls since the positive impact dominates. If the variation
continues to rise and becomes very large, the miss rate begins to
rise, as the need to raise the threshold helps Eve and counteracts
the positive impact. When $\sigma_ T$ becomes so large that both
the fixed part of the channel response and the thermal noise are
relatively negligible (i.e., $\sigma ^2 _T \gg \sigma ^2 _N$,
$\sigma ^2 _T \gg
{\sum_{m=1}^M|\overline{H}_{E,m}-\overline{H}_{A,m}|^2} $), then
using (\ref{E_beta_1}) we can rewrite the miss rate as
\begin{align}
\beta \approx \textrm{F}_{\chi^2_{2M}}((1-a)
\textrm{F}^{-1}_{\chi^2_{2M}}(1-\alpha)), \label{E_beta_3}
\end{align}
which is a function of the time-correlation of the temporal
variation parameter ($a$), frequency sample size ($M$), and the
false alarm rate ($\alpha$). If the variation is strongly
correlated in time ($a \approx 1$), the miss rate can be less than
that for the noise-dominated case, (\ref{E_beta_2}), where the
thermal noise is usually not negligible due to the limited
transmit power. An illustration of this trend will be given later.

Finally, we consider the impact of the spatial correlation of time
variations. The miss rate with total spatial correlation,
(\ref{E_beta_end}), decreases with $\mu$ in a manner that is
proportional to the inverse of \textbf{R}, (\ref{E_R1}), and thus
rises with $\sigma_T$. Since a strong spatial correlation of the
time variation damages the spatial variability character of the
channel, which is the basis of our scheme, it will degrade the
system performance.
\section{Simulation Methodology}\label{sec_sim}
\subsection{Simulating the Transfer Functions}
In order to test the proposed scheme, it is necessary to model not
only ``typical" channel responses, but the spatial variability of
these responses. Only in this way can we discern the success in
detecting would-be intruders like Eve. To that end, we make use of
the WiSE Tool, a ray-tracing software package developed by Bell
Laboratories \cite{WiSE:original}. One input to WiSE is the
3-dimensional plan of a specific building, including walls,
floors, ceilings and their material properties. With this
information, WiSE can predict the rays at any receiver from any
transmitter, including their amplitudes, phases and delays. From
this, it is straightforward to construct the transmit-receive
frequency response over any specified frequency interval
(bandwidth).

\begin{figure}[t]
\begin{center}
\epsfig{figure=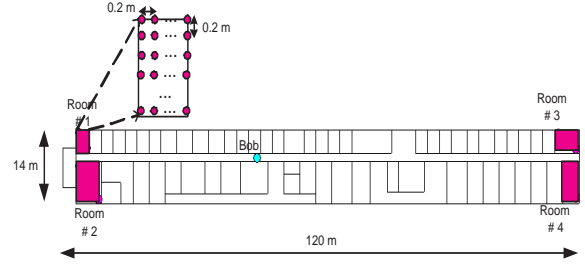, width=3in, height=1.4in}
\caption{System topology assumed in the simulations. Bob is
located at 2-m height near the center of a 120 m $\times$ 14 m
$\times$ 4 m office building. Alice and Eve are located on dense
grids at a height of 2 m. The sizes of the grids are $N_s=150$,
713, 315, and 348, respectively, for Room \# 1, 2, 3 and
4.}\label{fig_building}
\end{center}
\end{figure}

We have done this for one particular office building (the
Alcatel-Lucent Crawford Hill Laboratory in Holmdel, NJ), for which
a top view of the first floor is shown in Fig. \ref{fig_building}.
This floor of this building is 120 meters long, 14 meters wide and
4 meters high. For our numerical experiment, we placed Bob in the
hallway (the filled-in circle) at a height of 3 m. For the
positions of Alice and Eve, we considered four rooms at the
extremities of the building (shown shaded). For each room, we
assumed Alice and Eve both transmitted from a height of 2 m, each
of them being anywhere on a uniform horizontal grid of points with
0.2-meter separations. With $N_s$ grid points in a room, there
were $N_s(N_s - 1)/2$ possible pairs of Alice-Eve positions. For
Rooms 1, 2, 3 and 4, the numbers of grid points were $N_s = 150$,
713, 315 and 348, respectively. For each Alice-Eve pair, (1) WiSE
was used to generate the Alice-Bob and Eve-Bob average channel
responses ($\overline{{H}}_{A}(f)$ and $\overline{H}_{E}(f)$); and
(2) the hypothesis test described above was used to compute
$\beta$ for a specified $\alpha$. The set of all $\beta$-values in
a room were used to compute a room-specific mean,
$\overline{\beta}$, for each of several selected combinations of
bandwidth ($W$), number of frequency-domain samples ($M$),
transmit power ($P_T$), and channel variation models.

\subsection{Transmit Power, Receiver Noise, and Time Variation Strength}
Assume that, in conjunction with WiSE, we obtain the various
transfer functions as dimensionless ratios (e.g., received
\emph{E}-field/transmitted \emph{E}-field). Then the proper
treatment of the noise variance, $\sigma_N ^2$, in the hypothesis
test is to define it as the receiver noise power per tone, $P_N$,
divided by the transmit power per tone, $P_T/M$, where $P_T$ is
the total transmit power. Noting that $P_N  = \kappa \cT N_Fb$,
where $\kappa \cT$ is the thermal noise density in mW/Hz, $N_F$ is
the receiver noise figure, and $b$ is the measurement noise
bandwidth per tone in Hz \cite{RappaportBook}, we can write
\begin{align}
\sigma_N ^2 =\frac{\kappa \cT N_Fb}{P_T/M}=\frac{M}{\Gamma},
\label{E_sigma_N}
\end{align}
where $P_T$ is in mW, and $\Gamma = P_T/P_N$. We will henceforth
refer to $\Gamma$ by its decibel value.

Let $b^2_T$ denote the ratio between $\sigma_T ^2$ and the value
of $|H|^2$ averaged over the $M$ frequency samples (or ``tones")
and the $N_s$ receiver locations. We can thus write the standard
deviation of the time variation as
\begin{align}
\sigma_T =b_T \overline{\overline{H}}=b_T
\sqrt{\frac{1}{MN_s}\sum_{m=1}^M\sum_{l=1}^{N_s}|H_{l,m}|^2},
\end{align}
where $\overline{\overline{H}}$ can be regarded as a room
parameter, and $b_T$ represents the relative magnitude of the time
variation in a given room.

\section{Numerical Results}\label{sec_results}
In our simulations, we set $f_0=5$ GHz, $N_F=10$ (10 dB noise
figure), $\kappa \cT = 10^{-17.4}$ mW/Hz, $b=0.25$ MHz, $a=0.9$,
and, unless specified otherwise, $\alpha=0.01$
\cite{IEEE_TEE:LINKBUDGET}. As noted earlier, we place Alice and
Eve on dense grids in each of four rooms at the corners of a
particular building, with Bob in the hallway, Fig.
\ref{fig_building}. We obtained a miss rate for each Alice-Eve
pair in each room, and then calculated the average mean value for
each room in the building. Among them, Room \# 4, as the farthest
room from Bob, is likely to have the poorest performance in
rejecting Eve. In that sense, it lower-bounds the capabilities of
our PHY-layer authentication algorithm. For reasons of space, we
will only present results for Room \# 4, keeping in mind that they
are essentially worst-case or close to it.

\begin{figure}[t]
\begin{center}
\epsfig{figure=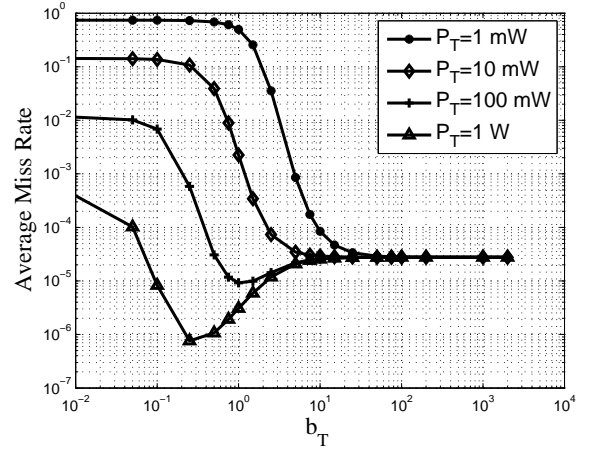, width=3.3in} \caption{The
average miss rate as function of the relative standard deviation
of the time variation, for the channel with spatially independent
temporal variation. $M=10$, $W=10$ MHz, $a=0.9$ and
$B_c=0$.}\label{fig_result_time_variance}
\end{center}
\end{figure}

Figure \ref{fig_result_time_variance} confirms the efficacy of the
algorithm in the presence of channel time variations. We assume
realistic system parameter values ($P_T=1$ mW $\sim$ 1 W, $M=10$
and $W=10$ MHz), and find that most average miss rates are smaller
than 0.01. The per tone signal-to-noise ratio (SNR) in the channel
measurements ranges from -12.8 dB to 14.2 dB, with a media value
of 6.4 dB, if using $P_T=10$ mW, $M=10$ and $W=10$ MHz. Also, as
pointed out in Section \ref{sec_test}, our proposed algorithm can
exploit the time variations to improve performance. For example,
the miss rate falls from around 0.01 to $10^{-5}$ when $b_T$ rises
from $0.01$ to $1$, with $P_T=100$ mW. The trend of these curves
with time variation confirms the discussion in Section
\ref{subsec:variation}, e.g., the minimum average miss rate is a
tradeoff between the positive impact of the time variation and its
negative impact resulting from the rise of the threshold.
Moreover, the miss rate falls with the transmit power $P_T$, as
expected, since it reduces the measurement noise at the receiver.

\begin{figure}[t]
\begin{center}
\epsfig{figure=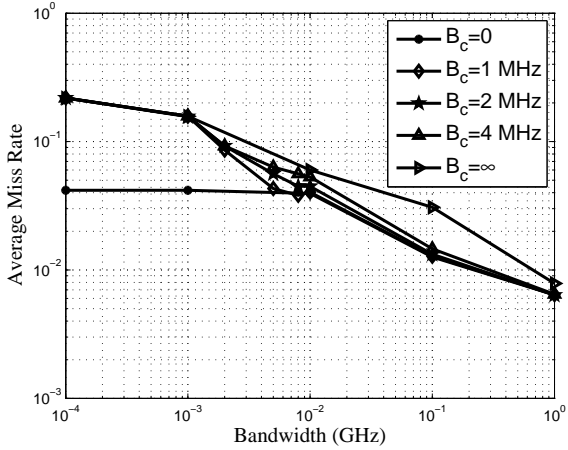, width=3.3in} \caption{The
average miss rate as a function of the measurement bandwith $W$,
for the channel with spatially independent temporal variation.
$M=5$, $a=0.9$, $b_T=0.5$, and $P_T=10$ mW.}\label{fig_result_W}
\end{center}
\end{figure}

Figure \ref{fig_result_W} demonstrates the impact of the bandwidth
$W$ and the coherence bandwidth $B_c$. We note that the results
for $B_c=0$ and $B_c=\infty$ are the lower and upper bounds,
respectively, of the miss rates, as well as the asymptotic results
for the high- and low-bandwidth regions. It is clear that
frequency correlations degrade performance. A related finding is
that the miss rate decreases with increasing bandwidth, $W$, since
the frequency response samples are more independent with larger
$W$.

\begin{figure}[t]
\begin{center}
\epsfig{figure=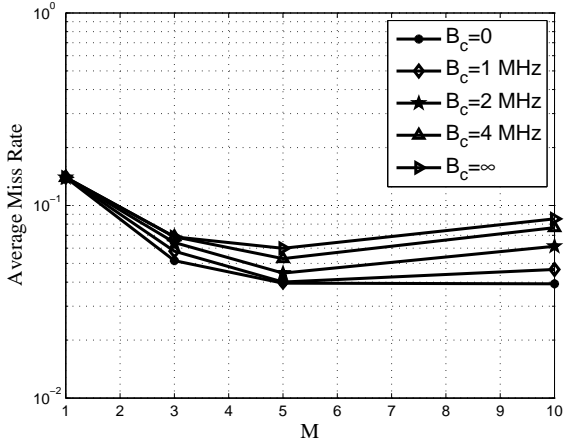, width=3.3in} \caption{The
average miss rate as a function of the number of frequency
samples, for the channel with spatially independent temporal
variation. $W=10$ MHz, $a=0.9$, $b_T=0.5$, and $P_T=10$
mW.}\label{fig_result_M}
\end{center}
\end{figure}

Figure \ref{fig_result_M} indicates that there is little benefit
(or even a deficit) in increasing $M$ beyond $\sim 10$, unless the
frequency correlation is very small (e.g., $B_c=0$) with high
transmit power. Actually, the optimal sample size $M$ in terms of
miss rate for specified measurement bandwidth decreases with the
coherence bandwidth $B_c$, because the noise power
(\ref{E_sigma_N}) rises with $M$ and the frequency-response
samples are more correlated with larger $B_c$.

\begin{figure}[t]
\begin{center}
\epsfig{figure=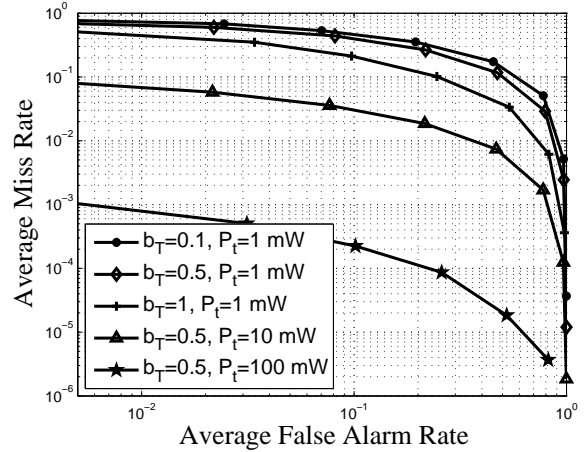, width=3.3 in}
\caption{Average miss rate vs. average false alarm rate when Bob
does not know the channel parameters, for the channel with
spatially independent temporal variation. $W=50$ MHz, $M=10$,
$a=0.9$, and $B_c=2$ MHz.}\label{fig_result_UNKNOWN}
\end{center}
\end{figure}

We see in Fig. \ref{fig_result_UNKNOWN} that the algorithm works
even when Bob does not know the key channel parameters, although
it requires either more transmit power or greater tolerance for
Type II errors. Interestingly, the time-variation may still help
here, e.g., the miss rate falls as $b_T$ rises from 0.1 to 1. How
to set the test threshold ${\mathcal T}$ in this case is an open
topic.

\begin{figure}[t]
\begin{center}
\epsfig{figure=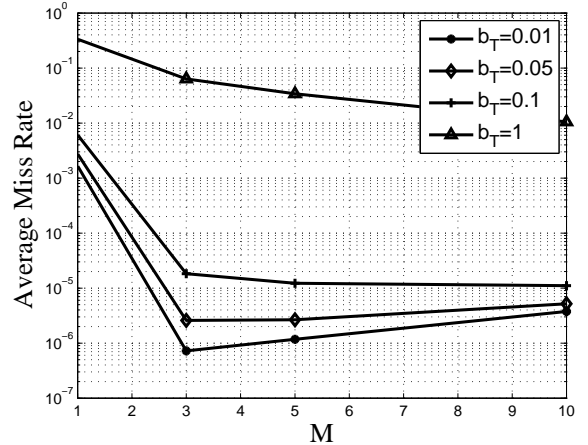, width=3.3 in}
\caption{Average miss rate vs. $M$, for the channel with totally
spatially correlated temporal variation. $W=100$ MHz, $B_c=2$ MHz,
$P_T=0.5$ W, and $a=0.9$. }\label{fig_result_spatial_correlation}
\end{center}
\end{figure}

Finally, Fig. \ref{fig_result_spatial_correlation} shows that the
system is very sensitive to the time-variation in an extreme case
of full spatial correlation. It requires much more transmit power
($P_T \sim 0.5$W) to reach the same miss rate performance. The
reason is quite simple: the mechanism of our scheme is to utilize
the spatial variability of the channel responses. The spatial
correlation of the time variation decreases the overall spatial
variability and thus degrades the performance.

The above assertions apply as well to the other shaded rooms in
Fig. \ref{fig_building} and, we can safely assume, to the other
rooms in the building.

\section{Conclusion}\label{sec_conclusion}
We have described and studied a physical layer technique for
enhancing authentication in a time-variant wireless environment.
Specifically, we assume the user terminals are stationary but
changes in the environment produce additive time-varying changes
in the channel responses. The technique uses channel frequency
response measurements and hypothesis testing to discriminate
between a legitimate user (Alice) and a would-be intruder (Eve).
With the ability to utilize the temporal-variation, it works even
when the receiver does not know the key channel variation
parameters, namely, the AR temporal coefficient $a$, the coherence
bandwidth $B_c$ and the standard deviation of the variation
$\sigma_T$, although these parameters help reduce the miss rate if
known.

The algorithm has been verified in a typical in-building
environments, where we used the ray-tracing tool WiSE to generate
realistic average channel responses and used a multipath tapped
delay line channel model for the temporal variation part of the
channel response. Simulation results have confirmed the efficacy
of the algorithm for realistic values of the measurement bandwidth
(e.g., $W \sim 10$ MHz), number of response samples (e.g., $M \leq
10$) and transmit power (e.g., $P_T >10$ mW). The miss rate is
generally smaller than 0.01, for a specified false alarm rate of
0.01, in the presence of moderate channel time variations.

We have found that the channel time variations can improve the
performance, e.g., the miss rate falls from around 0.01 to
$10^{-5}$ when the variation index $b_T$ rises from $0.01$ to $1$,
with $P_T=100$ mW. In addition, the miss rate decreases with the
transmit power of the probing signal and the measurement
bandwidth, and usually requires frequency samples of fewer than
10. We have also shown that the time correlation of the channel
variation is helpful, while coherence in the frequency and spatial
domains are harmful.

Research is currently in progress to address the case of user
terminal mobility. Effort is also needed, for the stationary case,
to explore the parameter space (e.g., the temporal coherence term
$a$); devise means of setting the test threshold ${\mathcal T}$;
consider other buildings; and conduct experiments to more
accurately characterize the time-variation properties of indoor
channels. Moreover, we are working to integrate physical layer
authentication into a holistic cross-layer framework for wireless
security that will augment traditional ``higher-layer" network
security mechanisms with physical layer methods.

 \appendices
\section{Proof of Relationship 1}

Since $A_l[k] \sim CN(0,P_\tau [l])$, from Eq. (\ref{e_epsil}) and
(\ref{e_var_al}), we have
\begin{align}
E[\epsilon_{A,m}[k]]=\sum_{l=0}^{\infty}E[A_l[k]e^{-j2\pi
(f_o-W/2+m \Delta f)l \Delta}]=0
\end{align}
and
\begin{align}
\textrm{Var}[\epsilon_{A,m}[k]]&=\sum_{l=0}^{\infty}\textrm{Var}[A_l[k]e^{-j2\pi
(f_o-W/2+m \Delta f)l
\Delta}]\nonumber\\&=\sum_{l=0}^{\infty}\textrm{Var}[A_l[k]]\nonumber\\&=\sum_{l=0}^{\infty}\sigma
_T^2(1-e^{-\gamma\Delta\tau})e^{- \gamma \Delta \tau l}=\sigma_T^2
\label{e_epsil_var}
\end{align}
Here we also utilize the fact that any two different multipath
components in a WSSUS channel are uncorrelated, i.e., $\forall l_1
\neq l_2$, $\forall k_1,k_2$,
\begin{align}
E[A_{l_1}[k_1]A_{l_2}[k_2]]=0 \label{e_us}
\end{align}

Considering that $A_l[k-1]$ and $u_l[k]$ are both zero-mean and
independent to each other, we see from Eq. (\ref{e_var_al}) and
(\ref{e_at}) that
\begin{align}
E[A_{l}[k_1]A_{l}[k_2]]&=a^{|k_1-k_2|}\textrm{Var}[A_{l}[\min
(k_1, k_2)]]\nonumber\\&=a^{|k_1-k_2|}\sigma_T^2
(1-e^{-\gamma\Delta\tau})e^{-\gamma\Delta\tau}\label{eq_E_AA}
\end{align}

Then from (\ref{e_epsil}), (\ref{e_us}), and (\ref{eq_E_AA}), we
have
\begin{align}
E&[\epsilon_{A,m}[k_1]\epsilon_{A,n}[k_2]^*]\nonumber\\
&=\sum_{l_1=0}^{\infty}\sum_{l_2=0}^{\infty} E[A_{l_1}[k_1]A_{l_2}[k_2]] \nonumber\\& \quad \cdot e^{-j2\pi[(f_o-W/2+m \Delta f){l_1}-(f_o-W/2+n \Delta f){l_2}]/W}\nonumber\\
&=\sum_{l=0}^{\infty}E[A_{l}[k_1]A_{l}[k_2]]e^{j2\pi (n-m)\Delta f
l/W}\nonumber\\
&=\sum_{l=0}^{\infty}a^{|k_1-k_2|}\sigma_T^2
(1-e^{-\gamma\Delta\tau})e^{- \gamma \Delta \tau l} e^{j2\pi
(n-m)\Delta f l/W}
\nonumber\\&=\frac{a^{|k_1-k_2|}\sigma_T^2(1-e^{-\gamma\Delta\tau})}{1-e^{-\gamma\Delta\tau+j2\pi
(n-m)\Delta f l \Delta\tau}}\label{eq_E_epsi3}
\end{align}

By (\ref{eq_chA}), (\ref{e_epsil_var}), and (\ref{eq_E_epsi3}), we
get
\begin{align}
& \textrm{Var}[H _{A,m}[k]-H _{A,m}[k-1]]\nonumber\\&=
\textrm{Var}[\epsilon _{A,m}[k]-\epsilon
_{A,m}[k-1]+N_{A,m}[k]-N_{A,m}[k-1]]\nonumber\\
&=\textrm{Var}[\epsilon _{A,m}[k]]+\textrm{Var}[\epsilon
_{A,m}[k-1]]\nonumber\\&\quad -2\textrm{Cov}[\epsilon
_{A,m}[k],\epsilon _{A,m}[k-1]
]+\textrm{Var}[N_{A,m}[k]]\nonumber\\&\quad +\textrm{Var}[N_{A,m}[k-1]]\nonumber\\
&=2\sigma ^2 _T-2E[\epsilon _{A,m}[k]\epsilon^*
_{A,m}[k-1]]+2\sigma_N^2\nonumber\\&=2\sigma^2_T(1-a)+2\sigma_N^2
\end{align}
The thermal noise components are independent of each other and all
the other variables, and the fixed part of the channel can be
viewed as constant, so $\forall m \neq n$
\begin{align}
&r(m-n)\nonumber\\&= \textrm{Cov}[H _{A,m}[k]-H _{A,m}[k-1],H _{A,n}[k] -H _{A,n}[k-1]]\nonumber\\
&= \textrm{Cov}[\epsilon _{A,m}[k]-\epsilon _{A,m}[k-1],\epsilon
_{A,n}[k]-\epsilon
_{A,n}[k-1]]\nonumber\\
&=E[\epsilon _{A,m}[k]\epsilon ^*_{A,n}[k]]+E[\epsilon
_{A,m}[k-1]\epsilon ^*_{A,n}[k-1]]\nonumber\\&\quad -E[\epsilon
_{A,m}[k]\epsilon
^*_{A,n}[k-1]]-E[\epsilon _{A,m}[k-1]\epsilon ^*_{A,n}[k]]\nonumber\\
&=\frac{2\sigma^2_T(1-a)(1-e^{-\gamma \Delta \tau})}{1-e^{-\gamma
\Delta \tau+j2\pi (n-m)\Delta \tau \Delta
f}}\nonumber\\&=\frac{2\sigma^2_T(1-a)(1-e^{-2\pi
B_c/W})}{1-e^{-2\pi B_c/W+j2\pi (n-m)/M}}
\end{align}

It can be easily proved that $\textbf{H} _{A}[k]-\textbf{H}
_{A}[k-1]$ has zero mean and is a Gaussian random variable, since
it is the linear combination of Gaussian random variables.

\section{Proof of Relationship 2}

For the case with spatially independent time variation where
$\epsilon_{E,m}[k]$ and $\epsilon_{A,m}[k]$ are independent
identically distributed, from (\ref{eq_chA}) and (\ref{eq_chE1})
we have
\begin{align}  \textrm{Var}&[H
_{E,m}[k]-H _{A,m}[k-1]]\nonumber\\&= \textrm{Var}[\epsilon
_{E,m}[k]-\epsilon
_{A,m}[k-1]+N_{E,m}[k]-N_{A,m}[k-1]]\nonumber\\
&=\textrm{Var}[\epsilon _{E,m}[k]]+\textrm{Var}[\epsilon
_{A,m}[k-1]]\nonumber\\&\quad
+\textrm{Var}[N_{A,m}[k]]+\textrm{Var}[N_{E,m}[k]]\nonumber\\&=2\sigma^2_T+2\sigma_N^2
\end{align}
And $\forall m \neq n$,
\begin{align}
& \textrm{Cov}[H _{E,m}[k]-H _{A,m}[k-1],H _{E,n}[k]-H
_{A,n}[k-1]]\nonumber\\&= \textrm{Cov}[\epsilon _{E,m}[k]-\epsilon
_{A,m}[k-1],\epsilon _{E,n}[k]-\epsilon
_{A,n}[k-1]]\nonumber\\&=E[\epsilon _{E,m}[k]\epsilon
^*_{E,n}[k]]+E[\epsilon _{A,m}[k-1]\epsilon
^*_{A,n}[k-1]]\nonumber\\&=\frac{2\sigma^2_T(1-e^{-\gamma \Delta
\tau})}{1-e^{-\gamma \Delta \tau+j2\pi (n-m)\Delta \tau \Delta
f}}=r(m-n)/(1-a)
\end{align}
From (\ref{eq_chA}) and (\ref{eq_chE1}) we also see that
$E[\textbf{H} _{E}[k]-\textbf{H} _{A}[k-1]
]=\overline{\textbf{H}}_E-\overline{\textbf{H}}_A$. The other part
is similar to that of Relationship 1.

\end{document}